\documentclass[aps,prl,twocolumn,showpacs,showkeys,reprint]{revtex4-1}
\usepackage{graphicx}
\usepackage{epstopdf}
\usepackage{epsfig}

\usepackage{hyperref}

\DeclareGraphicsExtensions{.eps,.tif}

\usepackage[centertags]{amsmath}
\usepackage{nomencl}
\usepackage{amsfonts}
\usepackage{latexsym}
\usepackage{color}
\definecolor{DarkOlive}{rgb}{0.1047,0.2412,0.0064}
\definecolor{FireBrick}{rgb}{0.5812,0.0074,0.0083}
\definecolor{RoyalBlue}{rgb}{0.0236,0.0894,0.6179}
\definecolor{RoyalGreen}{rgb}{0.0236,0.6179,0.0894}
\definecolor{RoyalRed}{rgb}{0.6179,0.0236,0.0894}
\definecolor{LightBlue}{rgb}{0.8544,0.9511,1.0000}
\definecolor{Black}{rgb}{0.0,0.0,0.0}
\definecolor{FuncColor}{rgb}{1.0,0.0,0.0}
\usepackage{mathrsfs}
\usepackage{subfigure}

\newcommand{\ro}[1]{\left( {#1}\right)}

\newcommand{\GammaRange}{$10^3 \lesssim \Gamma \lesssim 10^4${}}

\begin{document}
  \title{Ideal Gas Behavior of a Strongly-Coupled Complex (Dusty) Plasma}
  \author{Neil P.\ Oxtoby}
  \altaffiliation[Current address: ]{Dept.~of Computer Science, University College London, Gower Street, \mbox{WC1E 6BT}, London, UK}
  \author{Elias J.\ Griffith}
  \author{C\'eline Durniak}
  \author{Jason F.\ Ralph}
  \author{Dmitry Samsonov}
  \affiliation{Department of Electrical Engineering and Electronics, University of Liverpool, Liverpool, \mbox{L69 3GJ}, United Kingdom}
  \date{1 July, 2013}
  \begin{abstract}
    In a laboratory, a two-dimensional complex (dusty) plasma consists of a low-density ionized gas containing a confined suspension of Yukawa-coupled plastic microspheres.  For an initial crystal-like form, we report ideal gas behavior in this strongly-coupled system during shock-wave experiments.  This evidence supports the use of the ideal gas law as the equation of state for soft crystals such as those formed by dusty plasmas.
  \end{abstract}
  \maketitle
  
  An equation of state, such as the ideal gas law, is a mathematical relation between physical constants and macroscopically observable properties of a single phase of a system in equilibrium~\cite{LandauLifshitzStatMech}.  Equations of state are path-independent, and so can be explored by changing a system along any convenient intraphase path in state space between equilibria.  Interphase paths include a phase transition --- a discontinuous change in one or more system properties.  For example, the significant volume increase when liquid water evaporates.  Non-equilibrium paths, whether intraphase or interphase, can also be used to infer an equation of state, but an assumption is required to link the non-equilibrium states to the equilibrium states.  This is the case in shock-wave physics where otherwise unreachable high pressure and high density regions of state space are explored.  Pressure--density curves from shock-wave experiments do not provide enough thermodynamic information to infer an equation of state~\cite{Cowperthwaite1965} {(because other state variables also vary)}, but can be used to fit parameters in an assumed equation of state.  We explore parameter estimation in the ideal gas equation of state, applied to a two-dimensional complex plasma.  We demonstrate that this strongly-coupled system can be described by the ideal gas law, which is strictly valid only for systems of weakly-interacting particles.
  
  A laboratory complex plasma consists of plastic microspheres suspended in a low-density ionized gas.  The microspheres are often referred to as \textit{dust} particles in analogy with dusty plasmas observed in astronomy~\cite{Merlino2004,Newbury1997}.  Fast-moving electrons and relatively slow-moving ions in the plasma deposit a net negative charge on the dust, which repel each other via a screened Coulomb force (Yukawa or Debye-H\"uckel) \cite{Shukla2009:RMP}.  Condensed-matter-like behavior results when the dust is confined electrostatically, with the dust mimicking microscopic constituents of a fluid (individual molecules or atoms), yet being observable on a macroscopic scale (even to the naked eye).  The space between dust particles is occupied by a rarefied gas, so these dusty plasma structures experience weak damping, and are therefore considered to be representative models of liquids and solids~\cite{Morfill2009}.  Dusty plasmas are an excellent vehicle for exploring the microscopic kinematics of melting processes and crystal formation.  These kinematics are influenced by the local coupling constant $\Gamma$, which is the ratio of (interparticle interaction) potential energy to (thermal) kinetic energy for each particle.  Ideal gases are weakly coupled with $\Gamma<1$.
  
  A thermodynamic description of the dust is provided by {state variables} which can be calculated from the kinematics of the individual particles.  Individual particle positions extracted from images are used to determine both the dust density (via Voronoi analysis~\cite{Voronoi1908,Aurenhammer91voronoidiagrams}), and the coupling constant~\cite{Knapek2007}.  Particle velocities are used to determine the kinetic temperature~\cite{Oxtoby2012a,Samsonov2008:IEEE}.  {Fluctuations in these statistical properties are negligible in the thermodynamic limit, and at equilibrium.  For finite systems out of equilibrium, the statistical description retains validity, but fluctuations will be non-negligible.}
  
  Dust kinematics are normally estimated using particle tracking velocimetry (PTV)~\cite{Stegeman1995PTV}, where \textit{average} velocity $\vec{v}_\mathrm{PTV}(t+T/2)=[\vec{x}(t+T)-\vec{x}(t)]/T$ is calculated from consecutive position measurements $\vec{x}(t+T)$, $\vec{x}(t)$, which are extracted from a sequence of images taken with a high-speed camera at a frame rate of $1/T$ (typically 500--1000 frames per second).  The velocity calculated in this way is subject to two sources of inaccuracy: position uncertainty in the measurement, and nonzero acceleration.  For very high frame rates $T\rightarrow 0$, $v_\mathrm{PTV}$ is limited by position uncertainty, which is due to finite pixel size and noise in the camera sensor~\cite{Feng2007,FengRSI11}.  These limitations can lead to artifacts in results calculated from PTV-estimated kinematics.  Recursive state estimation (also known as \textit{object tracking}~\cite{BarShalom,JFRopaedia}) has been employed to estimate the kinematics of dusty plasma particles~\cite{hadziavdic2006,Oxtoby2012a}.  Object tracking algorithms filter noisy measurements via a set of equations to produce estimates of the \textit{instantaneous} kinematics which are resilient to the limitations discussed above.  The most ubiquitous recursive Bayesian estimator is the Kalman filter~\cite{Kalman1960}.
  
  In this work we employed object tracking using an interacting multiple model tracker~\cite{Oxtoby2012a} based on Kalman filtering (KF) to generate thousands of simultaneous particle tracks from shock-wave experiments on a two-dimensional (2D) dusty plasma.  We used Rankine-Hugoniot relations~\cite{BondGasDynamics} to calculate Hugoniot curves arising from the estimated kinematics, and observed ideal gas behavior despite the strong coupling between the dust particles ($\Gamma\gg1$).  Our experimental data fit the combined ideal gas/Rankine-Hugoniot model very well, but more complex models may be necessary for other regions of parameter space.  We compared our KF results with those from PTV.  The PTV results are unreliable due primarily to the significant particle acceleration in shock-wave experiments, and we observed a resulting systematic error that gave rise to a bias in the parameter estimation.  Our object tracking algorithm avoids this bias by including particle acceleration, along with position and velocity, in the recursive estimation.
  
  The ideal gas law is a thermodynamic relation between state variables.  It can be written in terms of specific (per unit mass) pressure $p$, internal energy $e$ and density $n$ as
    \begin{equation}
      p(e,n) = (\gamma - 1) e n ~,
      \label{eq:EoS_ideal_gas}
    \end{equation}
    where $\gamma$ is the adiabatic index.  Strictly speaking, the ideal gas law is a valid description for systems of non-interacting particles, but it can be applied to systems involving non-negligible particle interactions with sufficient accuracy in many cases~\cite{Pandey2004}.  Deviations from the ideal gas law were first considered by van der Waals~\cite{VanDerWaals1873} to account for finite particle size and interactions.
    
\begin{figure}[ht] 
    {\centering
    \includegraphics[width=0.9\columnwidth]{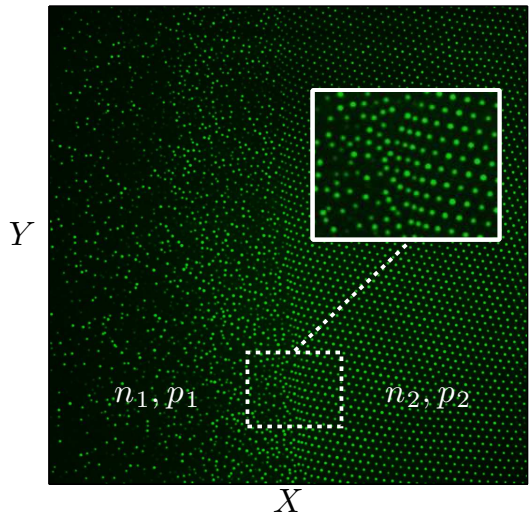} 
    \caption{Enhanced experimental image (enlarged dots, false color) with zoomed inset.  The field of view is 32.8mm/1024 pixels square.  Number density $n_{1,2}$ and specific pressure $p_{1,2}$ show the downstream and upstream regions (subscript 1/2).}
    \label{fig:image}
    }
  \end{figure}

    Here we explore the $p(e,n)$ relation in a non-perturbative manner by {generating a series of normal shock waves} of different magnitudes in the dust~\cite{Samsonov2003,Samsonov2004,Samsonov2008:IEEE}.  A normal shock wave is one where the shock front is normal to the direction of propagation, and the bulk flow is one-dimensional.  In the frame of a normal shock wave moving at speed $u_\mathrm{S}$, the Rankine-Hugoniot jump relations for conservation of mass, momentum and energy across the shock front are, respectively,~\cite{BondGasDynamics}
    \begin{subequations}
      \label{eqs:RankineHugoniot}
      \begin{align}
        \label{eq:RankineHugoniotMass_ShockFrame}
        n_2\tilde{u}_2 &= n_1\tilde{u}_1  \\
        \label{eq:RankineHugoniotMomentum_ShockFrame}
        \tilde{p}_2 + n_2\tilde{u}_2^2 &= \tilde{p}_1 + n_1\tilde{u}_1^2 \\
        \label{eq:RankineHugoniotEnergy_ShockFrame}
        e_2 + \frac{1}{2}\tilde{u}_2^2 + \frac{\tilde{p}_2}{n_2} &= e_1 + \frac{1}{2}\tilde{u}_1^2 + \frac{\tilde{p}_1}{n_1} ~, 
      \end{align}
    \end{subequations}
    where $u=u_\mathrm{S}-\tilde{u}$ is particle speed in the laboratory frame, a tilde denotes the reference frame of the shock wave, and the downstream (upstream) region is denoted with subscript 1 (2) --- see Fig.~\ref{fig:image}.  Number density $n$ and specific internal energy $e$ are equal in the laboratory and moving frames, but pressure has a kinetic component.  Using the technique of~\cite{Knapek2007} we used the particle kinematics to find \GammaRange\ in the crystal state ahead of the shock wave, implying negligible kinetic pressure ($\tilde{p}_1~\approx~p_1$).  We observed a similar trend upstream, but the technique of~\cite{Knapek2007} cannot be applied in the wake of the shock wave due to the disorder, so we calculated $\tilde{p}_{1,2}$ in this work.  Equation~(\ref{eq:RankineHugoniotEnergy_ShockFrame}) is known as \textit{the Hugoniot}~\cite{Henderson2001}.  Using equation~(\ref{eq:EoS_ideal_gas}) to eliminate internal energy from equation~(\ref{eq:RankineHugoniotEnergy_ShockFrame}), and combining with equation~(\ref{eq:RankineHugoniotMomentum_ShockFrame}), we can write~\cite{BondGasDynamics}
    \begin{equation}
      \xi(\eta)
      = 
      \frac{\eta\ro{\gamma+1} - (\gamma-1)}{\ro{\gamma+1} - \eta(\gamma-1)} ~,
        \label{eq:HugoniotEoS_Final}
    \end{equation}
    where $\xi\equiv \tilde{p}_2/\tilde{p}_1$ is the shock strength and $\eta\equiv n_2/n_1$ is the compression ratio across the shock front.  An estimate of $\gamma$, and hence an approximate equation of state for the shocked dust in the form of equation~(\ref{eq:EoS_ideal_gas}), is obtained from a least-squares fit of the experimental results to equation~(\ref{eq:HugoniotEoS_Final}).
    
  With the ideal gas law as the equation of state, the polytropic index $g$ can be used to describe the physical nature of a process that changes an initial state (downstream $p_1, n_1$) to a final state (upstream $p_2, n_2$). Polytropic processes follow $p/n^g = C$~\cite{Newbury1997}, which is a curve in the pressure--density diagram, with $g$ and $C$ defining a solution for the changes linking initial and final states.  Equating initial and final states (both equal to the constant $C$) then combining with $\xi$ and $\eta$ and solving for $g$ allows the polytropic index to be expressed as
  \begin{equation}
    g =\frac{\ln(\xi)}{\ln(\eta)} ~,
    \label{eq:Polytropic_index}
  \end{equation}
  where $g=0$ indicates an isobaric process, $g=1$ is an isothermal process and $g=\gamma$ is an adiabatic process.
  
  The experiment involves levitating a 2D cloud of microspheres 10mm above the floor of an Argon-filled chamber pressurized to $2.05\mathrm{Pa}$.  The spheres (9.2$\mu$m diameter) were allowed to settle into a well-spaced crystalline structure, forming a ``plasma crystal"~\cite{Samsonov2004} which is visible to the naked eye when illuminated by a laser sheet (figure~\ref{fig:image}).  The dust particles each hold an approximate charge of $Q=16000e$ and have a Debye length of $\lambda_D=1.0\mathrm{mm}$~\cite{Harvey2010,Durniak2010:IEEE}.  Shock waves were created by an electrode located to one side of the field of view, which was pulsed for 2 seconds with a voltage selected from $-20$V to $-50$V in $5$V steps.  The crystal was allowed to reset between each run (requiring approximately 100 seconds).  The experiment was repeated at each voltage level to reduce the impact of local variation in crystal structure that can form on reset.  The dust was imaged from above by a grayscale camera at 500 frames per second for 1.2 seconds, and the resulting images processed by PTV and our Kalman-filter-based tracking algorithm to obtain the dust kinematics.  A sample image of the dust, enhanced for presentation with enlarged dots and false color, is shown in figure~\ref{fig:image} with a zoomed inset around the shock front.  Further details of the experimental apparatus are described in~\cite{Samsonov2004} and~\cite{Samsonov2005}.
  
  The symmetry inherent in normal shock waves permits a 1D description of the dynamics.  \textit{Profile} values were calculated as robust average quantities (median) in each of 50 bins which were equally spaced along the $X$ axis {(the direction of propagation)}, and which spanned the $Y$ axis.  Density and pressure profiles were used in our analysis.  Density is the inverse of the Voronoi cell area~\cite{Voronoi1908,Aurenhammer91voronoidiagrams}, and the pressure is normal stress (in the direction of propagation), which here is the 
first diagonal component of the 2D stress tensor, $P_{XX}$. 
  
\begin{figure}[ht] 
    {\centering
    \includegraphics[width=0.9\columnwidth]{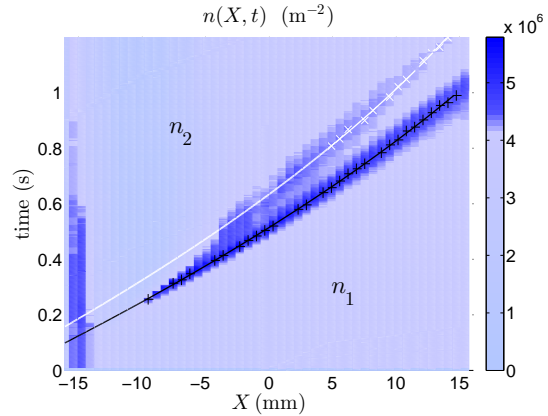}
    \caption{The dust number density profile evolution $n(X,t)$ showing the shock wave (black circles), and trailing wave (white squares) with quadratic least-squares fits.}
  \label{fig:time_evolution}
  }
  \end{figure}

\begin{figure}[ht] 
  {\centering
    \includegraphics[width=0.9\columnwidth]{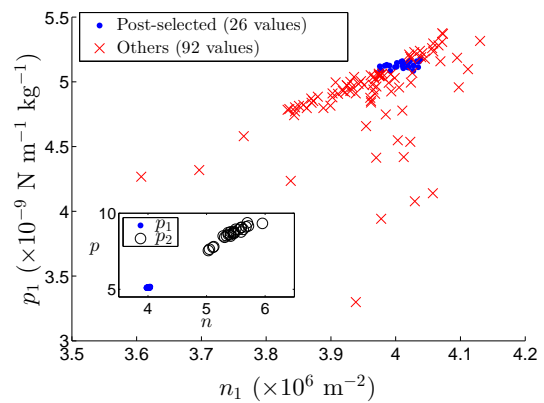}
    \caption{Initial pressure and density ($n_1,p_1$) for each run: blue dots survived post-selection (see text).  
    Inset: Pressure--density diagram showing all post-selected data: initial states (blue dots) and the corresponding final states (black circles).}
    \label{fig:selected_points}
  }
  \end{figure}

\begin{figure}[ht] 
    {\centering
      \includegraphics[width=0.9\columnwidth]{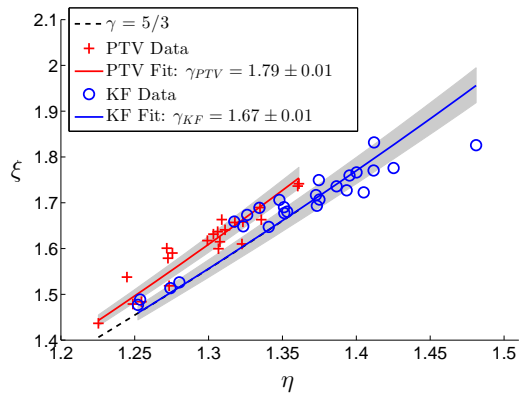}
      \caption{Shock strength vs.\ compression ratio for PTV and KF.  Least-squares fits to equation~(\ref{eq:HugoniotEoS_Final}) give the adiabatic index $\gamma$, with $3\sigma$ confidence regions in each case shown in gray.  The KF fit overlaps with that of a monatomic ideal gas.}
      \label{fig:xi_eta}
    }
  \end{figure}

  Our investigation proceeded as follows.  The shock front was identified as a peak in the density profile evolution (figure~\ref{fig:time_evolution}), from which the shock front position and speed was determined.  The upstream and downstream quantities in the Rankine-Hugoniot shock-jump relations (pressure, density, etc.)\ were selected from $0.656$mm (1 bin) behind the shock front and $3.28$mm (5 bins) ahead.  We needed to look further ahead to overcome the finite width of the shock front (an ideal shock wave would have vanishing width).  Results were also sensitive to the chosen upstream distance due to structure a few millimeters behind the shock wave (see multi-shock discussion below).  
Shock-wave (Hugoniot) investigations such as in this Letter require repeated shock-wave experiments of different magnitudes, sharing a common initial condition.  Reliably reproducing the same initial condition in dust crystal {experiments} is extremely difficult, if not impossible.  For this reason the data was post-selected from the densest cluster of initial conditions and constrained to lie within 1\% of the cluster centroid.  This is illustrated in figure \ref{fig:selected_points} where the post-selected initial conditions (downstream) are shown as blue dots and all others as red crosses.  The inset of figure \ref{fig:selected_points} shows the final states corresponding to the post-selected initial conditions.  From $13$ similar experimental runs, $118$ total data points were generated, of which $26$ were post-selected.  The apparently small dynamic range of the post-selected data in figure~\ref{fig:xi_eta} is typical for shock-wave experiments in dusty plasmas~\cite{Melzer1994,Samsonov2000,Samsonov2004,Durniak2010:IEEE}.  It is a consequence of the crystal softness (very strong shock waves completely destroy soft crystals), which is due to the large interparticle spacing relative to the particle size~\cite{FengSound2012}.  
The post-selected data was analyzed using equations 
(\ref{eqs:RankineHugoniot})--(\ref{eq:Polytropic_index}).
  
  {A typical experiment is visualized in} figure~\ref{fig:time_evolution}.  Two number density peaks emerged following the applied voltage pulse: a shock wave (black line) and a trailing wave (white line).  Such multi-shock structures~\cite{Duvall1977} can be described by a sequence of jump relations like equation (\ref{eqs:RankineHugoniot}).  Wave speeds calculated from least-squares fits for the peak positions were $u_\mathrm{S}(t) =  -17.2t + 43.7$ mm/s (shock, $t\geq0.24$s) and $u_\mathrm{T}(t) = -8.4 t + 33.0$ mm/s (trailing, $t\geq0.45$s).
  
  {The adiabatic index of the dust was estimated by least-squares fits to equation~(\ref{eq:HugoniotEoS_Final}).}  Figure~\ref{fig:xi_eta} shows these fits of shock strength vs.\ compression for both PTV and KF (object tracking).  We found $\gamma_\mathrm{KF} = 1.67~\pm~0.01$, which is consistent with that of a monatomic ideal gas $\gamma_\mathrm{Ideal}=5/3=1.6\dot{6}$.  We found $\gamma_\mathrm{PTV} = 1.79~\pm~0.01$ using PTV.  This is a biased overestimate, as we now explain.  The PTV and KF results for the crystal-like downstream states were comparable, so the source of the PTV bias was the upstream estimates of $\tilde{p}_2$ and $n_2$.  Dust pressure is dominated by the Yukawa interaction, which is non-linear in interparticle spacing $r$ (e.g., see~\cite{Oxtoby2012a}), and so sensitive to errors in $r$.  These errors are greater for PTV than KF~\cite{Oxtoby2011Fusion,Oxtoby2012a} and, when averaged, propagate through the non-linearities to create a biased \textit{overestimate} of upstream pressure.  This shifts erroneous results upward in the $\xi$--$\eta$ plane.  Dust density is \textit{underestimated} when shocked dust particles intermittently leave the plane of illumination, shifting erroneous results left in the $\xi$--$\eta$ plane.  Object tracking provides a robust way to maintain tracks for these particles, whereas PTV does not.  Thus, as observed in figure~\ref{fig:xi_eta}, we expected the PTV result to lie above and to the left of the KF result (which itself would lie above the true result if biases were present).
  
  We determined the polytropic index of the shocked dust via the mean value of equation~(\ref{eq:Polytropic_index}), using KF results.  We found $g_\mathrm{KF} = 1.71~\pm~0.07$ which satisfies $g \approx \gamma$, thereby demonstrating that shock waves in a dusty plasma crystal constitute an adiabatic process, as is the case for an ideal gas~\cite{Newbury1997}.  This is further experimental evidence of ideal gas behavior in a 2D dusty plasma. 
  
\begin{figure}[htp] 
    {\centering
      \includegraphics[width=0.9\columnwidth]{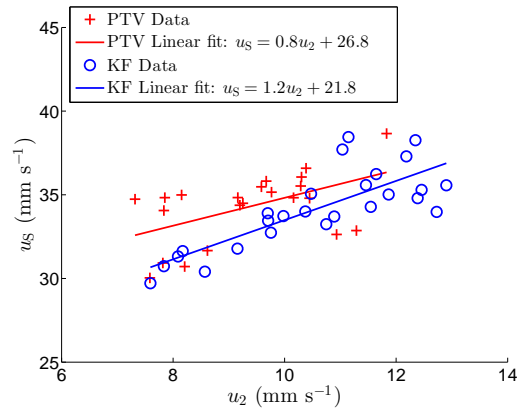}
      \caption{Shock front speed vs.\ upstream particle speed (\textit{Shock Hugoniot}) for PTV and KF.}
      \label{fig:shock_hugoniot}
    }
  \end{figure}

  Our final result is the shock Hugoniot~\cite{Rice1958,Nagayama2002} in figure~\ref{fig:shock_hugoniot}, where shock wave speed $u_\mathrm{S}$ is linearly related to upstream particle speed $u_2$: $u_\mathrm{S} = S u_2 + C_0$.  Here $C_0$ is the zero-pressure bulk speed of sound (for an unshocked sample), and $S$ is a dimensionless constant of proportionality for the linear fit.  The PTV and KF methods estimate $C_0$ to be $26.8$mm/s and $21.8$mm/s, respectively.  These values are in line with the $25$mm/s and $28$mm/s speeds of sound observed in \cite{FengSound2012} and \cite{Schwabe2011} via different techniques.  The very low speeds result from the extreme softness of the dust crystal.  For the fits in figure~\ref{fig:shock_hugoniot}, the coefficient of determination $R^2$ showed the KF data ($R^2=0.64$) following the expected linear trend far better than the PTV data ($R^2 = 0.27$).  This reinforces our conviction that object tracking methods should be used to analyze dusty plasma experiments, rather than PTV.  

  In this work we performed shock-wave experiments on a 2D dusty plasma crystal, a system of strongly-coupled particles with Coulomb coupling parameter $\Gamma \sim 10^3$.  We calculated state variables for the dust (pressure, density) directly from the dust particle kinematics.  The kinematics were estimated using two techniques: object tracking (recursive Bayesian state estimation), and particle tracking velocimetry (the standard approach in dusty plasma physics, which is less accurate~\cite{Oxtoby2012a}, and unreliable for shock-wave experiments).  Conservation laws (Rankine-Hugoniot equations) were combined with the ideal gas law to estimate the adiabatic index of the dust, which revealed a significant finding: a strongly-coupled ($\Gamma \gg 1$) 2D dusty plasma behaves as an ideal gas.  This is explained by the relatively low compression ratio tolerable by soft crystals, e.g.~dusty plasma crystals, which negates the need for higher-order density terms as found in equations of state for non-ideal gases.  While the ideal gas law combined with the Rankine-Hugoniot equations produced a very good fit to the experimental data, more complex models may be required when accessing different regions of parameter space.
  
 We acknowledge financial support from the Engineering and Physical Sciences Research Council of the United Kingdom (grant number EP/G007918) and high-throughput computational resources provided by the eScience team at the University of Liverpool.  We thank the anonymous referees for helpful critiques.  The experiments were performed by D.S., who passed away during the final preparation stages of this paper.  Dmitry will be sorely missed by the plasma physics community.

\bibliography{DustyPlasmas}

\end{document}